\documentclass[11pt]{article}
\usepackage{graphicx}
\usepackage{amssymb}
\usepackage{hyperref}
\begin{document}

\def\ch{{\rm ch}}
\def\sh{{\rm sh}}
\def\ba{\begin{eqnarray}}
\def\ea{\end{eqnarray}}
\def\nn{\nonumber}

\def\A{Ai\left(-\frac{E}{F^{2/3}}\right)}
\def\B{Bi\left(-\frac{E}{F^{2/3}}\right)}

\def\dA{Ai^\prime\left(-\frac{E}{F^{2/3}}\right)}
\def\dB{Bi^\prime\left(-\frac{E}{F^{2/3}}\right)}

\def\Ana{Ai\left(-\frac{(E-F a)}{F^{2/3}}\right)}
\def\Bna{Bi\left(-\frac{(E-F a)}{F^{2/3}}\right)}
\def\dAna{Ai^\prime\left(-\frac{(E-F a)}{F^{2/3}}\right)}
\def\dBna{Bi^\prime\left(-\frac{(E-F a)}{F^{2/3}}\right)}

\def\Apa{Ai\left(-\frac{(E+F a)}{F^{2/3}}\right)}
\def\Bpa{Bi\left(-\frac{(E+F a)}{F^{2/3}}\right)}
\def\dApa{Ai^\prime\left(-\frac{(E+F a)}{F^{2/3}}\right)}
\def\dBpa{Bi^\prime\left(-\frac{(E+F a)}{F^{2/3}}\right)}

\newtheorem{theorem}{Theorem}
\newtheorem{corollary}[theorem]{Corollary}
\newtheorem{definition}{Definition}

\title{Simple Soluble Molecular Ionization Model\footnote{Work towards a 
UConn Senior Thesis (CG).}}
\author{Gerald V. Dunne and Christopher S. Gauthier\\
Department of Physics\\
University of Connecticut\\
Storrs, CT 06269-3046}
\date{}
\maketitle

\begin{abstract}
We present a simple exact analytical solution, using the Weyl-Titchmarsh-Kodaira 
spectral theorem, for the spectral function of the one-dimensional diatomic 
molecule model consisting of two attractive delta function wells in the 
presence of a static external electric field. For sufficiently deep and far 
apart wells, this molecule supports both an even and an odd state, and the 
introduction of a static electric field turns these bound states into 
quasi-bound states which are Stark shifted and broadened. The continuum 
spectrum also inherits an intricate pattern of resonances  which reflect the 
competition between resonant scattering between the two atomic wells and 
between the linear potential and one or both atomic well(s). All results are 
analytic and can be easily plotted. The relation between the large orders of 
the divergent perturbative Stark shift series and the non-perturbative 
widths of quasi-bound levels is studied.
\end{abstract}

\section{Introduction}
\label{intro}

The electronic structure of atoms and molecules is usefully probed by external 
electric and magnetic fields. The development of intense lasers has 
permitted such probing in regimes where simple perturbative treatments are 
not valid and one must use a nonperturbative semiclassical approximation or 
a numerical approach. Experiments with molecules display a much richer range 
of phenomena than with atoms, due to the additional molecular degrees of 
freedom \cite{codling,bucksbaum,nibarger}. These include above threshold ionization 
\cite{corkum,hansch}, multiple ionization \cite{walker,bandrauk}, alignment 
effects \cite{dietrich}, electron localization \cite{misha1}, non-sequential 
double ionization \cite{talebpour}, direct excitation \cite{gibsonpra}, 
stabilization \cite{schmidt}, dissociative recombination \cite{talebpour2}, 
and separation effects \cite{barnett}. However, molecules (even the simplest 
diatomic molecules) and their ionization are clearly more difficult to treat 
theoretically. Many approximation techniques have been developed and applied 
to atomic ionization processes \cite{froelich,geltman,reinhardt,yamabe,adk}, 
but much less is known for molecular systems. Realistic calculations are 
rather complicated and one loses some of the physical intuition that can 
often be gained from simple models. In this paper we present the exact 
analytical solution for a simple molecular ionization model. The molecule is 
taken to be one-dimensional. This approximation is remarkably good in the 
strong-field regime where the ionization is predominantly along the field 
direction, so that the system is effectively one-dimensional 
\cite{dietrich}. The simplest such one-dimensional molecule consists of two 
atomic wells\footnote{It is straightforward to generalize the exact solution 
to the case of unequal atomic well depths, but for simplicity here we 
consider the atomic well depths to be equal.} represented by attractive 
delta function wells of strength $g$, separated by a distance $2a$. This is 
a well-known soluble model \cite{gas}. This molecule always supports an even 
parity bound ground state and a continuum, and if $ag>1$ it also supports an 
odd parity bound excited state. There is a long tradition of using model 
potentials such as zero-range potentials in atomic and molecular physics 
\cite{demkov}.
The presence of an external electrostatic field, of field strength $F$, 
dramatically changes the basic character of the spectrum, converting the 
bound states into quasi-bound states and modifying the resonance structure 
of the continuum. These spectral changes are seen directly in the spectral 
function $\rho(E)$ : the quasi-bound states are poles of $\rho(E)$ at 
complex values of the energy $E$, where the real part of the pole gives the 
energy location of the quasi-bound level and the imaginary part gives the 
width, and hence lifetime, of the level. Since this molecular ionization 
model is exactly soluble, we can easily investigate the dependence of these 
quasi-bound levels on the relevant physical parameters -- the field strength 
$F$, the atomic well depth $g$, and the atomic separation parameter $a$. The 
same applies for the "continuum", where resonance structures appear due to 
the delicate interplay between tunneling, binding and scattering effects.

The solution presented here uses the Weyl-Titchmarsh-Kodaira (WTK) spectral 
theorem \cite{weyl,titchmarsh,kodaira,richtmyer}. This spectral theorem 
expresses the completeness of the wavefunctions of the Schr\"odinger 
equation in a general way that applies not just to the familiar discrete 
spectrum models (such as the infinite square well or the harmonic 
oscillator), but also to systems with discrete and continuum spectra (such 
as the finite square well or the hydrogen atom), and even to systems with a 
purely continuous spectrum, such as for ionization problems where there are 
no true bound states. The WTK approach is well suited for numerical 
implementation and has been applied long ago to the Stark effect in atomic 
hydrogen \cite{hydrogenweyl}. An interesting soluble one-dimensional atomic 
model consisting of a single finite square well on the half-line is solved 
using the WTK method in \cite{dean}. More recently, the numerical WTK 
approach has been used for studying strong-field ionization effects in 
effectively one-dimensional diatomic molecules \cite{gibson,barnett}. 
Various numerical and approximate methods for  computing resonance locations 
and widths are compared  in \cite{sidky}. This current paper is 
complementary to \cite{gibson}, but the choice to represent the atomic wells 
by delta function wells makes the entire molecular ionization problem 
analytically solvable, thereby bypassing the numerical part of the 
computation. We also mention that this one-dimensional molecular ionization
model has been studied in \cite{glasser} 
using the solution of the associated Lippman-Schwinger equation.

This paper is organized as follows. Section 2 contains a summary of the 
implementation of the Weyl-Titchmarsh-Kodaira (WTK) method for computing the 
spectral function. This is applied in Section 3 to the one-dimensional 
molecular model without an applied electric field. In Section 4 the electric 
field is applied and the exact solution for the spectral function is 
derived. The dependence of the spectral function on the various physical 
parameters is explored through plots and also analytically. Section 5 is 
devoted to the simpler case of atomic ionization, obtained from the 
molecular solution by taking the atomic separation parameter, $a$, to zero. 
In this case we also study the large orders of perturbation theory for the 
Stark shift and show its relation to the nonperturbative level width. The 
final Section contains some concluding comments.

\section{Weyl-Titchmarsh-Kodaira Method}
\label{wtk}

The Weyl-Titchmarsh-Kodaira (WTK) spectral theorem 
\cite{weyl,titchmarsh,kodaira,richtmyer} for quantum mechanical Hamiltonians 
is very general, covering not just simple Hamiltonians like the harmonic 
oscillator which have only bound states, but also Hamiltonians with both 
bound and continuum states. It also extends to ionization problems where the 
spectrum is purely continuum. Indeed, in his classic book \cite{titchmarsh}, 
Titchmarsh solves the half-line "atomic" problem of a binding delta function 
potential well plus a constant electric field. This approach can be applied 
directly to any one-dimensional or radial Schr\"odinger problem. The WTK 
method can be summarized as follows.

Consider the Schr\"odinger equation (we work in units where 
$\frac{\hbar^2}{2m}=1$):
\ba
-\psi^{\prime\prime}(x) +V(x) \psi = E\psi(x)
\label{schrod}
\ea
on the real line $x\in (-\infty,+\infty)$. Pick some point, chosen without 
loss of generality to be $x=0$, and normalize the two independent solutions 
of (\ref{schrod}), $u(x,E)$ and $v(x,E)$, so that their Wronskian is equal 
to 1 at that point by choosing:
\ba
u(0,E)=1\quad &,& \quad u^\prime(0,E)=0 \nn\\
v(0,E)=0\quad &,& \quad v^\prime(0,E)=-1
\label{norm}
\ea
Next, integrate (numerically or analytically)  each of these solutions out 
towards $x=+\infty$ and $x=-\infty$, producing four functions  $u_\pm(x,E)$, 
and $v_\pm(x,E)$. The WTK method involves constructing particular linear 
combinations of these functions such that these combinations are 
normalizable on the intervals $(0, +\infty)$ and $(-\infty, 0)$, when the 
energy $E$ has a small positive imaginary part: $E\to E+i\epsilon$. 
Specifically, going towards the right, we construct
\ba
\psi_+ (x,E)=u_+(x,E)+m_+(E) v_+(x,E)
\label{mp}
\ea
such that it is normalizable on the interval $(0, +\infty)$ when the energy 
$E$ has a small positive imaginary part. This determines the coefficient 
function $m_+(E)$. Similarly, going to the left, we construct the linear 
combination
\ba
\psi_- (x,E)=u_-(x,E)+m_-(E) v_-(x,E)
\label{mm}
\ea
such that it is normalizable on the interval $(-\infty, 0)$ when the energy 
$E$ has a small positive imaginary part. This determines the coefficient 
function $m_-(E)$. The spectral function, and the completeness of the 
wavefunctions, can be expressed in terms of these coefficient functions 
$m_\pm(E)$ \cite{weyl,titchmarsh,kodaira,richtmyer}. For example, the spectral function is
\ba
\rho(E)=\lim_{\epsilon\to 0} \frac{1}{\pi} {\it Im} 
\left(\frac{m_+(E+i\epsilon)m_-(E+i\epsilon)+1}{m_+(E+i\epsilon)-m_-(E+i\epsilon)}\right)
\label{spectral}
\ea
This WTK method is well suited to numerical implementation for arbitrary 
potential wells \cite{gibson}. In the models studied in this paper the 
situation is even simpler, since the independent solutions $u(x,E)$ and 
$v(x,E)$ are known in analytic form for all $x$; the "integration" process 
simply involves applying the correct continuity and discontinuity boundary 
conditions at the locations of the two delta function potentials. This will 
be shown in detail in Section \ref{molelectric}.

\section{One-Dimensional Molecular Model}
\label{mol}
We first review the model without the electric field. We choose the 
following simple double-delta-function potential to represent the diatomic 
molecule:
\ba
V(x)= - g\left[\delta(x+a)+\delta(x-a)\right]
\label{potential}
\ea
where $g>0$. This potential has two binding delta-function wells located at 
$x=\mp a$, and chosen for simplicity to have equal strength $-g$. This 
problem, without an external electric field, is a standard problem in 
quantum mechanics courses \cite{gas}. The potential (\ref{potential}) always 
supports a bound ground state which has even parity, and if $ga>1$ it also 
supports a bound excited state which has odd parity. In this section we 
present  the WTK solution as an introductory illustration of the method.

In the vicinity of $x=0$ the potential (\ref{potential}) vanishes. Thus,  
the two independent solutions, $u(x,E)$ and $v(x,E)$, of the Schr\"odinger 
equation (\ref{schrod}), which satisfy the normalization conditions 
(\ref{norm}) are
\ba
u(x,E)=\cos(\sqrt{E}\, x) \quad , \quad v(x,E)= -\frac{\sin(\sqrt{E}\, 
x)}{\sqrt{E}}.
\label{freeuv}
\ea
Integrating to the right, these solutions remain valid until we reach the 
right-hand
delta-function at $x=+a$, at which point we impose the standard boundary 
conditions \cite{gas}:
\ba
\psi(a-\epsilon)=\psi(a+\epsilon)\quad ,\quad \frac{d\psi}{d 
x}\bigg|_{a-\epsilon}^{a+\epsilon}=-g\,\psi(a)
\label{bcs}
\ea
These conditions determine the solutions for $x>a$ to be:
\ba
u_+(x,E)=A\,\cos(\sqrt{E}\, x) +B\, \frac{\sin(\sqrt{E}\, x)}{\sqrt{E}}\\
v_+(x,E)=C\,\cos(\sqrt{E}\, x) +D\, \frac{\sin(\sqrt{E}\, x)}{\sqrt{E}}
\label{jump}
\ea
where the coefficients are
\ba
A=1+\frac{g \sin(2a\sqrt{E})}{2\sqrt{E}}\quad , \quad B=-g 
\cos^2(a\sqrt{E})\\
C=-\frac{g}{E}\sin^2(a\sqrt{E}) \quad , \quad D=-1+\frac{g 
\sin(2a\sqrt{E})}{2\sqrt{E}}
\ea
\begin{figure}[ht]
\includegraphics{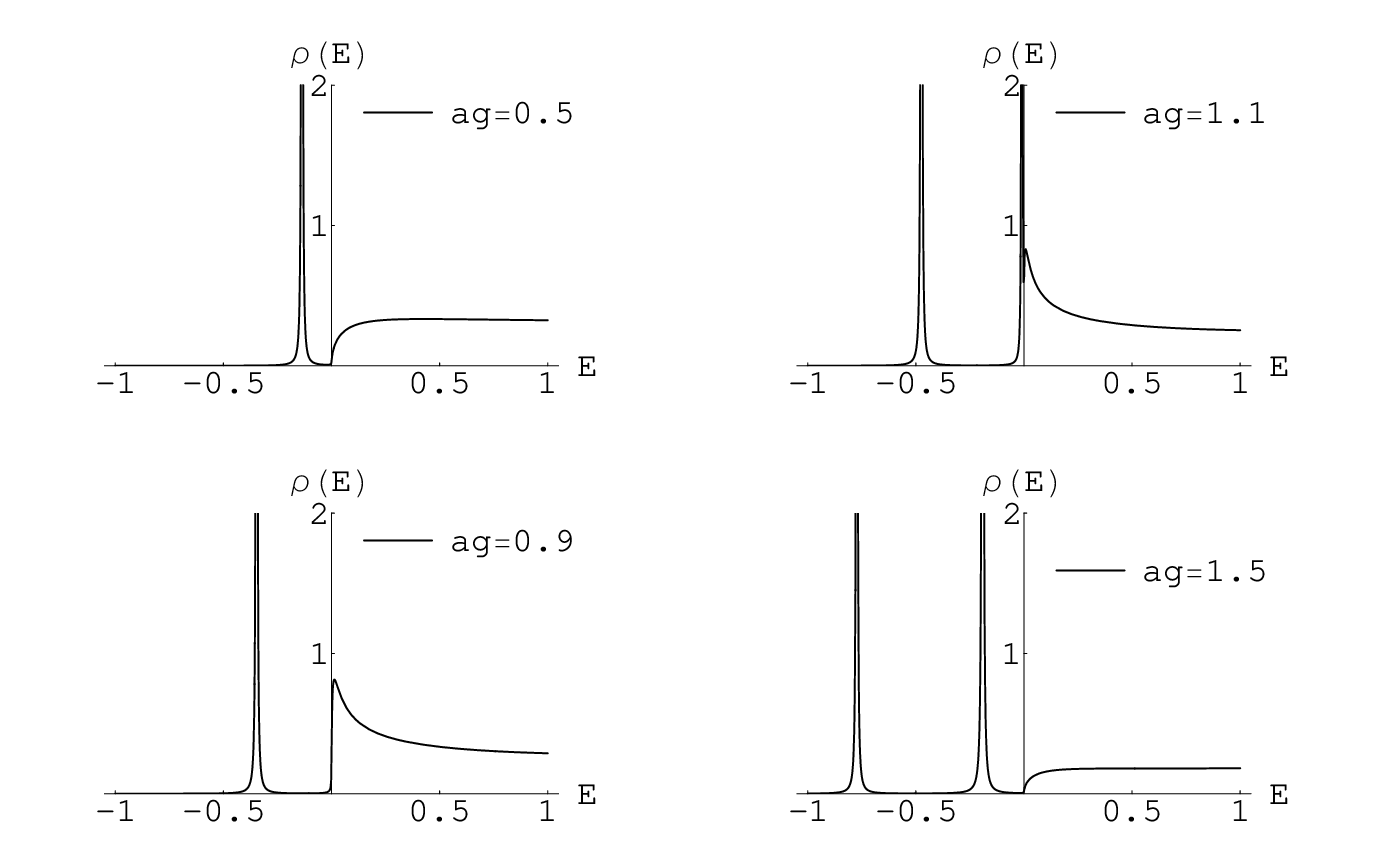}
\caption{Plots of the spectral function for the free one-dimensional 
molecular potential (\protect{\ref{potential}}). For $a g<1$ there is only 
one bound state, but for $a g>1$ there are two bound states.}
\label{freebound}
\end{figure}
When the energy $E$ has a small positive imaginary part, the linear 
combination $\psi_+=u_++m_+v_+ $ in (\ref{mp}) will be normalizable on $(0, 
+\infty)$ if the $\exp(-i\sqrt{E}x)$ part is eliminated. This determines the 
function $m_+(E)$ in (\ref{mp}) to be
\ba
m_+(E)&=& -\frac{\sqrt{E} A + i B}{\sqrt{E} C + i D}\nn\\
&=& \frac{-iE-g\sqrt{E}\cos(a\sqrt{E})\, 
e^{ia\sqrt{E}}}{\sqrt{E}-g\sin(a\sqrt{E})\, e^{ia\sqrt{E}}}
\ea
Since the potential (\ref{potential}) is symmetric, it follows that 
$m_-(E)=-m_+(E)$. Thus, the spectral function (\ref{spectral}) is determined 
to be:
\ba
\rho(E)=\lim_{\epsilon\to 0} \frac{1}{2\pi} {\it Im} \left( 
-m_+(E+i\epsilon)+\frac{1}{m_+(E+i\epsilon)}\right)
\label{molspectral}
\ea
Bound states appear on the negative real energy axis as poles of the 
spectral function (\ref{molspectral}). The pole of $m_+$, satisfying the 
transcendental equation
\ba
1+e^{-2a\sqrt{-E}}&=&\frac{2\sqrt{-E}}{g},
\label{moleven}
\ea
corresponds to the even parity bound state. From (\ref{molspectral}), the 
zero of $m_+$, satisfying the transcendental equation
\ba
1-e^{-2a\sqrt{-E}}&=&\frac{2\sqrt{-E}}{g},
\label{molodd}
\ea
also gives a pole of the spectral function, and corresponds to the odd 
parity bound state (if it exists).
There is always an even bound state, given by the pole of $m_+(E)$ 
satisfying (\ref{moleven}). If $ga>1$ there is also an odd bound state, 
given by the zero of $m_+(E)$ satisfying (\ref{molodd}).
\begin{figure}[ht]
\includegraphics{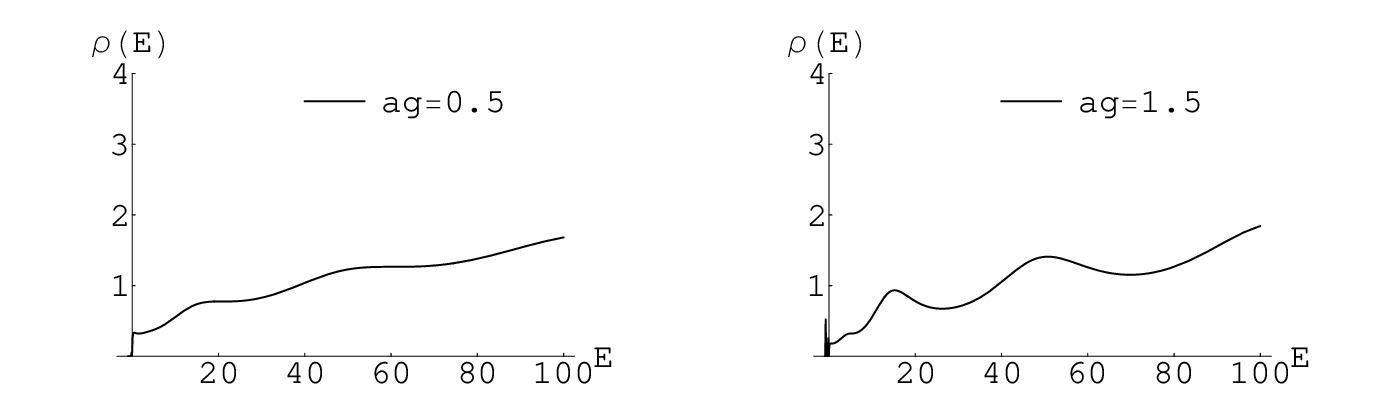}
\caption{Plots of the continuum part of the spectral function for the free 
one-dimensional molecular potential (\protect{\ref{potential}}). Note the 
periodic behavior due to resonant back-scattering between the two wells. 
The left-hand plot is for $ a g=0.5$, while the right-hand plot has $a 
g=1.5$.}
\label{freecontinuum}
\end{figure}
Some plots of the spectral function (\ref{molspectral}) are shown in Figs. 
\ref{freebound} and \ref{freecontinuum}. Note the appearance of the single 
bound state when $g a<1$, but of two bound states when $g a>1$. In Fig. 
\ref{freebound} the width of the bound state peaks in these plots is 
artificial, as we have kept the small imaginary part $\epsilon$ of the 
energy nonzero in order to show the peaks. In the true $\epsilon\to 0$ limit 
these bound state peaks have zero width, and so would not show up on the 
plot. Fig. \ref{freecontinuum} shows the continuum part of the spectrum -- 
notice the periodic behavior of the spectral function, due to resonant 
backscattering between the two delta-functions. This periodicity is 
determined by the separation $2a$ between the wells, and the well strength 
$g$.

\section{One Dimensional Molecular Model With an Electric Field}
\label{molelectric}

In this section we solve the problem of the one-dimensional molecular 
potential (\ref{potential}) studied in the previous section, with an 
additional external static electric field, of magnitude $F$. Thus, the 
potential is
\ba
V(x)=-g\left[\delta(x+a)+\delta(x-a)\right]-F\,x
\label{starkpotential}
\ea
where $g>0$, and $F>0$. The potential (\ref{starkpotential}) has no bound 
states when $F\neq 0$, but it does have quasi-bound states. The 
corresponding Schr\"odinger equation (\ref{schrod}) is analytically soluble 
since the linearly independent solutions are Airy functions. In fact, the 
WTK solution proceeds exactly as in the case without the electric field, 
except that the basic trigonometric solutions in (\ref{freeuv}) are replaced 
by Airy functions. 

\subsection{WTK Solution for the Spectral Function}
In the vicinity of $x=0$, the independent solutions satisfying the Wronskian 
normalization condition (\ref{norm}) are
\ba
u(x,E)&=&A^{(u)}\, Ai\left(-\frac{(F x+E)}{F^{2/3}}\right)+B^{(u)}\, 
Bi\left(-\frac{(F x+E)}{F^{2/3}}\right)
\nn\\
v(x,E)&=&A^{(v)}\, Ai\left(-\frac{(F x+E)}{F^{2/3}}\right)+B^{(v)}\, 
Bi\left(-\frac{(F x+E)}{F^{2/3}}\right)
\label{airyuv}
\ea
where $Ai$ and $Bi$ are Airy functions \cite{abram}, and the coefficients needed to satisfy the normalization conditions (\ref{norm})
are
\ba
A^{(u)}(E)=\pi\dB \quad &,&\quad B^{(u)}(E)=-\pi\dA\nn\\
A^{(v)}(E)=-\pi F^{-1/3} \B \quad &,&\quad B^{(v)}(E)=\pi F^{-1/3} \A
\ea
Here we have made use of the fundamental Airy function Wronskian identity 
\cite{abram}:
\ba
Ai(x)Bi^\prime(x)-Ai^\prime(x)Bi(x)=\frac{1}{\pi}
\quad , \quad  \forall \; x \in \mathbb{R}
\ea
As before, integrating to the right, the solutions in (\ref{airyuv}) remain 
valid until we reach the right-hand delta-function well at $x=+a$, at which 
point we apply the delta-function boundary conditions (\ref{bcs}). This 
determines the solutions in the region $x>a$ to be:
\ba
u_{+}(x,E)&=&A_{+}^{(u)}\, Ai\left(-\frac{(F 
x+E)}{F^{2/3}}\right)+B_{+}^{(u)}\, Bi\left(-\frac{(F 
x+E)}{F^{2/3}}\right)\nn\\
v_{+}(x,E)&=&A_{+}^{(v)}\, Ai\left(-\frac{(F 
x+E)}{F^{2/3}}\right)+B_{+}^{(v)}\, Bi\left(-\frac{(F x+E)}{F^{2/3}}\right)
\ea
where the coefficients are
\ba
A_{+}^{(u)}(E)&=& -g\pi^{2} F^{-1/3}\Apa \Bpa \dB
\nn\\
&& +g\pi^{2} F^{-1/3} \Bpa^{2} \dA+\pi \dB
\nn\\
B_{+}^{(u)}(E)&=& -g\pi^{2} F^{-1/3}\Apa \Bpa \dA
\nn\\
&& +g\pi^{2} F^{-1/3} \Apa^{2} \dB-\pi \dA
\nn\\
A_{+}^{(v)}(E)&=& g\pi^{2} F^{-2/3}\Apa \Bpa \B
\nn\\
&&-g\pi^{2} F^{-2/3} \Bpa^{2} \A-\pi F^{-1/3} \B
\nn\\
B_{+}^{(v)}(E)&=&
g\pi^{2} F^{-2/3}\Apa \Bpa \A
\nn\\
&&-g\pi^{2} F^{-2/3} \Apa^{2} \B+\pi F^{-1/3} \A
\label{pluscoeff}
\ea
Similarly, integrating to the left, the solutions (\ref{airyuv}) remain 
valid until we reach the left-hand delta-function well at $x=-a$, at which 
point we apply the boundary conditions (\ref{bcs}). This determines the 
solutions in the region $x<-a$ to be:
\ba
u_{-}(x,E)&=&A_{-}^{(u)}\, Ai\left(-\frac{(F 
x+E)}{F^{2/3}}\right)+B_{-}^{(u)}\, Bi\left(-\frac{(F x+E)}{F^{2/3}}\right)
\nn\\
v_{-}(x,E)&=&A_{-}^{(v)}\, Ai\left(-\frac{(F 
x+E)}{F^{2/3}}\right)+B_{-}^{(v)}\, Bi\left(-\frac{(F x+E)}{F^{2/3}}\right)
\ea
where the coefficients are
\ba
A_{-}^{(u)}(E)&=&g \pi^{2} F^{-1/3} \Bna \Ana \dB
\nn\\
&& -g\pi^{2} F^{-1/3} \Bna^{2} \dA+\pi \dB
\nn\\
B_{-}^{(u)}(E)&=& g \pi^{2} F^{-1/3} \Bna \Ana \dA
\nn\\
&& -g\pi^{2} F^{-1/3} \Ana^{2} \dB-\pi \dA
\nn\\
A_{-}^{(v)}(E)&=&-g \pi^{2} F^{-2/3} \Bna \Ana \B
\nn\\
&& +g\pi^{2} F^{-2/3} \Bna^{2} \A-\pi F^{-1/3} \B
\nn\\
B_{-}^{(v)}(E)&=&-g \pi^{2} F^{-2/3} \Bna \Ana \A
\nn\\
&& +g\pi^{2} F^{-2/3} \Ana^{2} \B+\pi F^{-1/3} \A
\label{minuscoeff}
\ea
Thus, the independent solutions to the Schr\"odinger equation for the 
potential (\ref{starkpotential}) are known analytically for all $x$. As 
described in Section \ref{wtk}, the WTK method involves finding the 
functions $m_\pm(E)$ such that the linear combinations $\psi_\pm=u_\pm+m_\pm 
v_\pm$ are normalizable as $x\to\pm \infty$, when the energy has a small 
positive imaginary part.

In the region $x>a$, the normalizable solution has the form
\ba
\psi_+(x,E)\propto Ai\left(-\frac{(F x+E)}{F^{2/3}}\right) -i\, 
Bi\left(-\frac{(F x+E)}{F^{2/3}}\right)\label{rightdecay}
\ea
This determines $m_+(E)$ to be
\ba
m_{+}(E)=-\frac{B_{+}^{(u)}(E)+i\;A_{+}^{(u)}(E)}{B_{+}^{(v)}(E)+i\;A_{+}^{(v)}(E)}
\label{mplus}
\ea
In the region $x<-a$, the normalizable solution has the form
\ba
\psi_-(x,E)\propto Ai\left(-\frac{(F x+E)}{F^{2/3}}\right)
\label{leftdecay}
\ea
This determines $m_-(E)$ to be
\ba
m_{-}(E)=-\frac{B_{-}^{(u)}(E)}{B_{-}^{(v)}(E)}
\label{mminus}
\ea
Given these expressions for $m_\pm(E)$, the spectral function is given by 
(\ref{spectral}), which we can write in terms of the coefficients 
(\ref{pluscoeff}) and (\ref{minuscoeff}) as
\ba
\rho(E)=\lim_{\epsilon \rightarrow 0}
\frac{1}{\pi} Im
\textstyle\left(\frac{B_{-}^{(u)}(E+i\epsilon)\left[B_{+}^{(u)}(E+i\epsilon)+iA_{+}^{(u)}(E+i\epsilon)\right]+
B_{-}^{(v)}(E+i\epsilon)\left[B_{+}^{(v)}(E+i\epsilon)+iA_{+}^{(v)}(E+i\epsilon)\right]}
{B_{-}^{(u)}(E+i\epsilon)\left[B_{+}^{(v)}(E+i\epsilon)+iA_{+}^{(v)}(E+i\epsilon)\right]-
B_{-}^{(v)}(E+i\epsilon)\left[B_{+}^{(u)}(E+i\epsilon)+iA_{+}^{(u)}(E+i\epsilon)\right]}\right)
\label{starkspectral}
\ea
This is an analytic expression for the exact spectral function for the 
potential (\ref{starkpotential}). In the remainder of this Section we 
discuss the physical properties of this spectral function, using plots and 
analytical methods.

\subsection{Plots of the Spectral Function}

Before discussing the analytic properties of the spectral function 
(\ref{starkspectral}), we present some plots which illustrate how the 
spectral function depends on the physical parameters, in order to develop 
some intuition for the physical processes involved. In addition, some animations showing how the spectral function changes, both in the quasi-bound and in the quasi-continuum region, as we vary the field strength $F$, the atomic separation parameter $a$, or the atomic well depth parameter $g$ can be found at \cite{animations}.

\subsubsection{Dependence on the electric field strength $F$.}
\begin{figure}[ht]
\includegraphics{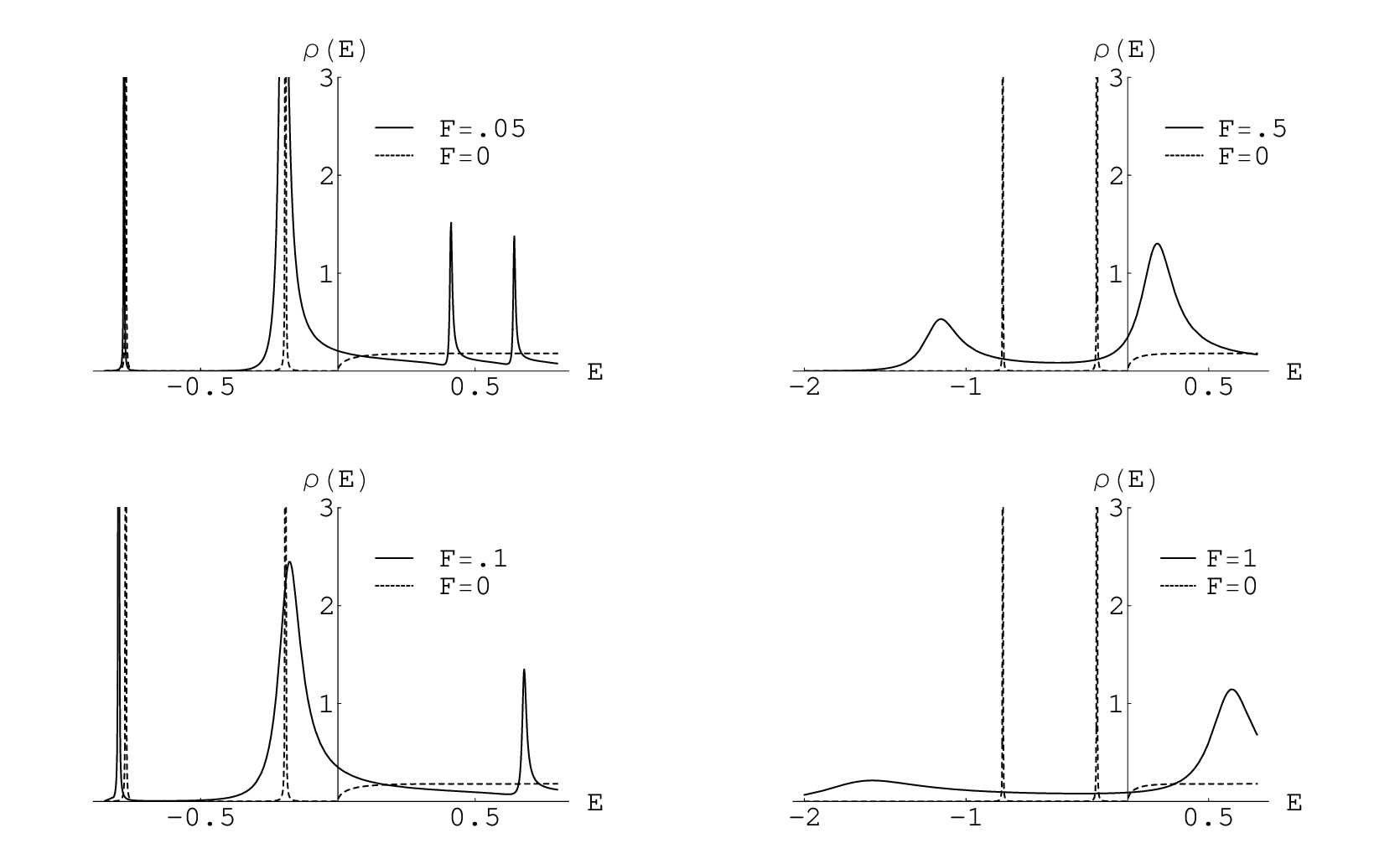}
\caption{Plots of the spectral function (\protect{\ref{starkspectral}}) 
[solid lines] illustrating how the quasi-bound states change as the 
electric field strength $F$ varies. The dashed line shows the corresponding 
$F=0$ molecular spectrum (\protect{\ref{molspectral}}. In these plots, the 
atomic well strength is $g=1.5$, and the separation parameter is $a=1$, so 
the free molecule has two bound states. Notice that as $F$ increases these 
two states are Stark shifted and broadened.}
\end{figure}
\begin{figure}[ht]
\includegraphics{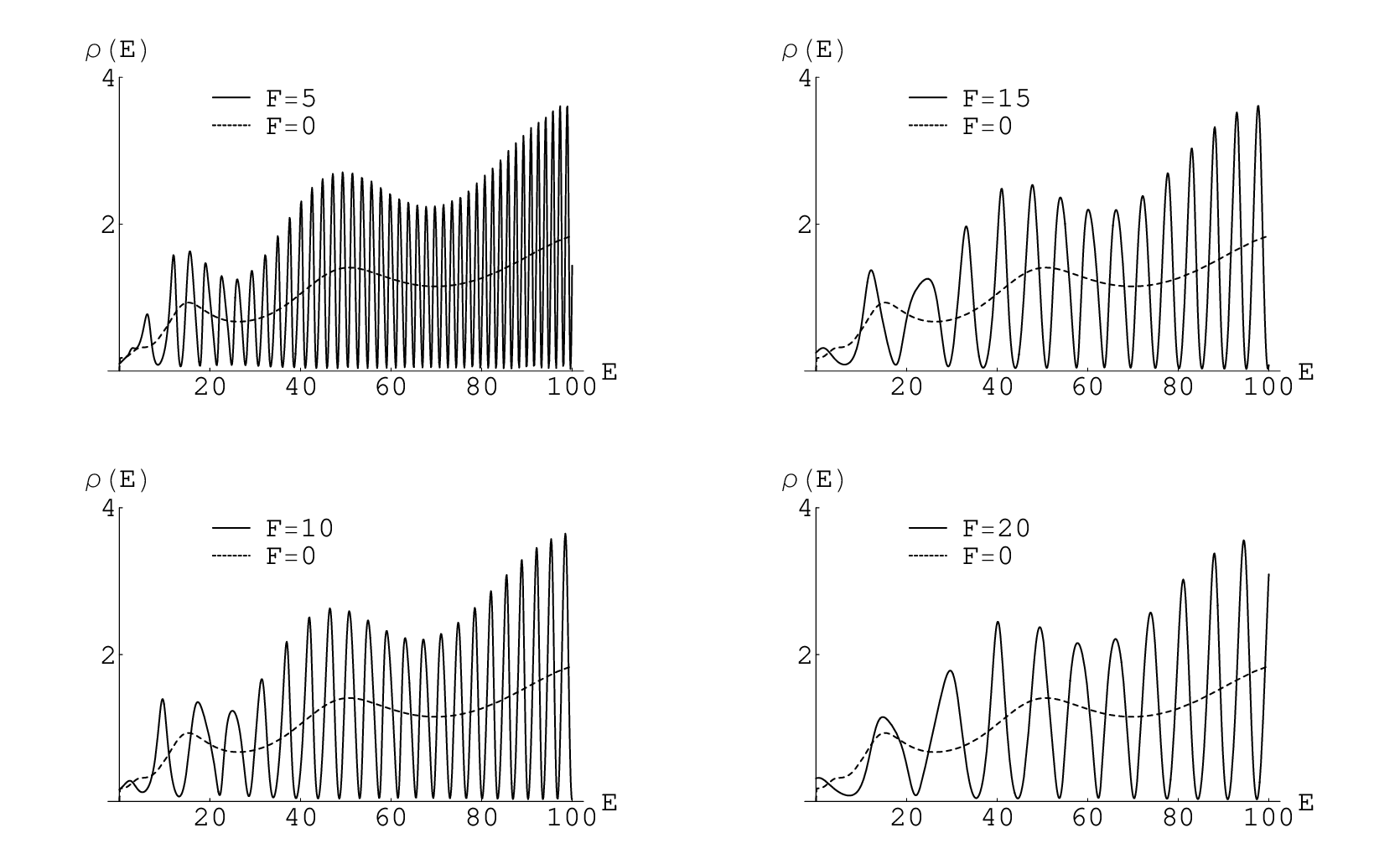}
\caption{Plots of the spectral function (\protect{\ref{starkspectral}}) 
[solid lines] illustrating how the continuum part of the spectrum changes 
as the electric field strength $F$ varies. The dashed line shows the 
corresponding $F=0$ molecular spectrum (\protect{\ref{molspectral}}). In 
these plots, the atomic well strength is $g=1.5$, and the separation 
parameter is $a=1$. Notice the two oscillation scales -- the longer one is 
set by the free case [dashed line], while the rapid oscillation is set by 
the field strength $F$, with
the free spectral function providing an average. As $F$ decreases in 
magnitude, the oscillations become more rapid, eventually averaging out to the 
free case.}
\end{figure}
In the absence of the electric field, the molecule has one or two bound 
states depending on the well separation parameter, $a$, and well strength, 
$g$. And the continuum exhibits structure due to Ramsauer-Townsend 
resonances in the scattering between the two delta wells. These two features 
are illustrated clearly in Figures 1 and 2. When the field strength $F$ is 
nonzero, the bound states become quasi-bound states with a nonzero width, 
and their central values are Stark shifted. These effects are illustrated in 
Figure 3, which shows a molecule with two bound states subjected to an 
external electric field of strengths $F=0.05, 0.1,0.5, 1$. Note that the 
quasi-bound states broaden as the field increases, with the higher state 
being broader since it is less deeply bound and so can tunnel more easily. 
The quasi-bound states are Stark shifted in opposite directions: the lower 
state is Stark shifted down in energy, while the higher state is Stark 
shifted up in energy.
\begin{figure}[ht]
\includegraphics{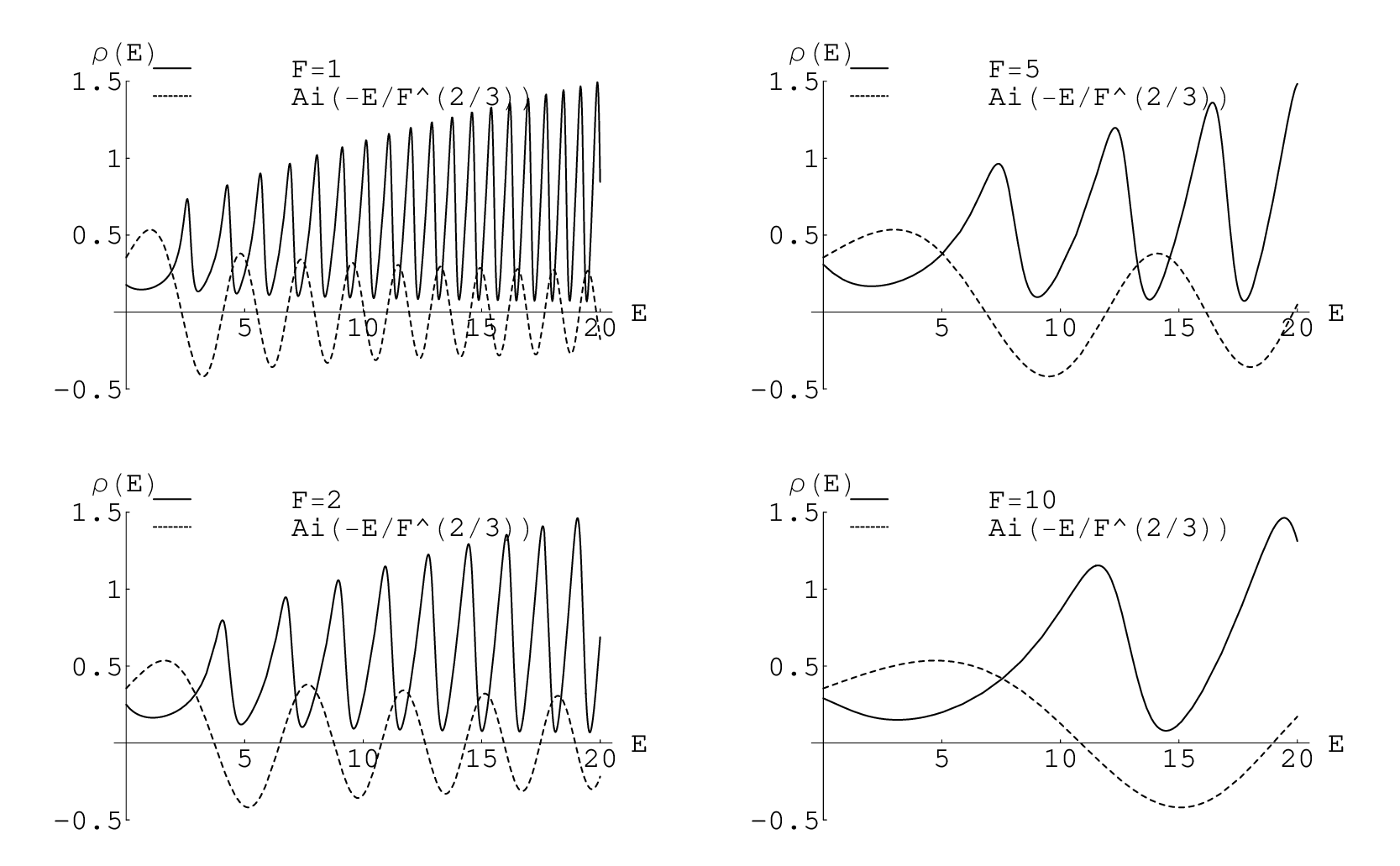}
\caption{Plots of the spectral function (\protect{\ref{starkspectral}}) 
[solid lines] illustrating how the continuum part of the spectrum 
correlates with the zeros of the Airy function $Ai(-E/F^{2/3})$ [dashed 
line], which are the energies of the wedge potential $V=-F x$ with an 
infinite wall at some point.  This is an illustration of the Ramsauer 
effect. These plots are for $a=0$ and $g=1.5$, and various values of the 
electric field strength $F$, as shown.}
\end{figure}
The effect of the external electrostatic field on the continuum states is 
shown in Figure 4. In these plots the dashed line shows the continuum 
spectral function when $F=0$. The solid lines show the spectral function for 
various values of $F$: $F=5, 10, 15, 20$. Note that the free spectral 
function provides an average for the $F>0$ spectral function. The scale of 
the oscillations of this average function is clearly independent of $F$, 
being determined by the combination $ag$, as illustrated in Figure 2. For 
$F>0$ there is an additional scale in the "continuum" spectrum, and as $F$ 
decreases to zero the rapid oscillations become more and more rapid, and 
eventually average out to the free spectral function. As $F$ increases 
the period of this oscillation increases. These oscillations can be 
correlated approximately with the zeros of the Airy function 
$Ai(-E/F^{2/3})$, since these zeros give the energies of the half-wedge 
potential well which has $V=-F x$ for $x<0$, with an infinite barrier at 
$x=0$. This is illustrated in Figures 5 and 6, for $a=0$ and $a=1$, 
respectively. Note that the peaks of the spectral function (the solid line) 
coincide roughly with the zeros of $Ai(-E/F^{2/3})$, the dashed line. The 
agreement is quite good, even for the molecular model having $a=1$. This is 
an example of Ramsauer-Townsend resonance, with the electron backscattering 
off the two delta wells providing the large period oscillations, and the 
electron scattering off the delta wells and the linear electrostatic 
potential providing the shorter period oscillations.
\begin{figure}[ht]
\includegraphics{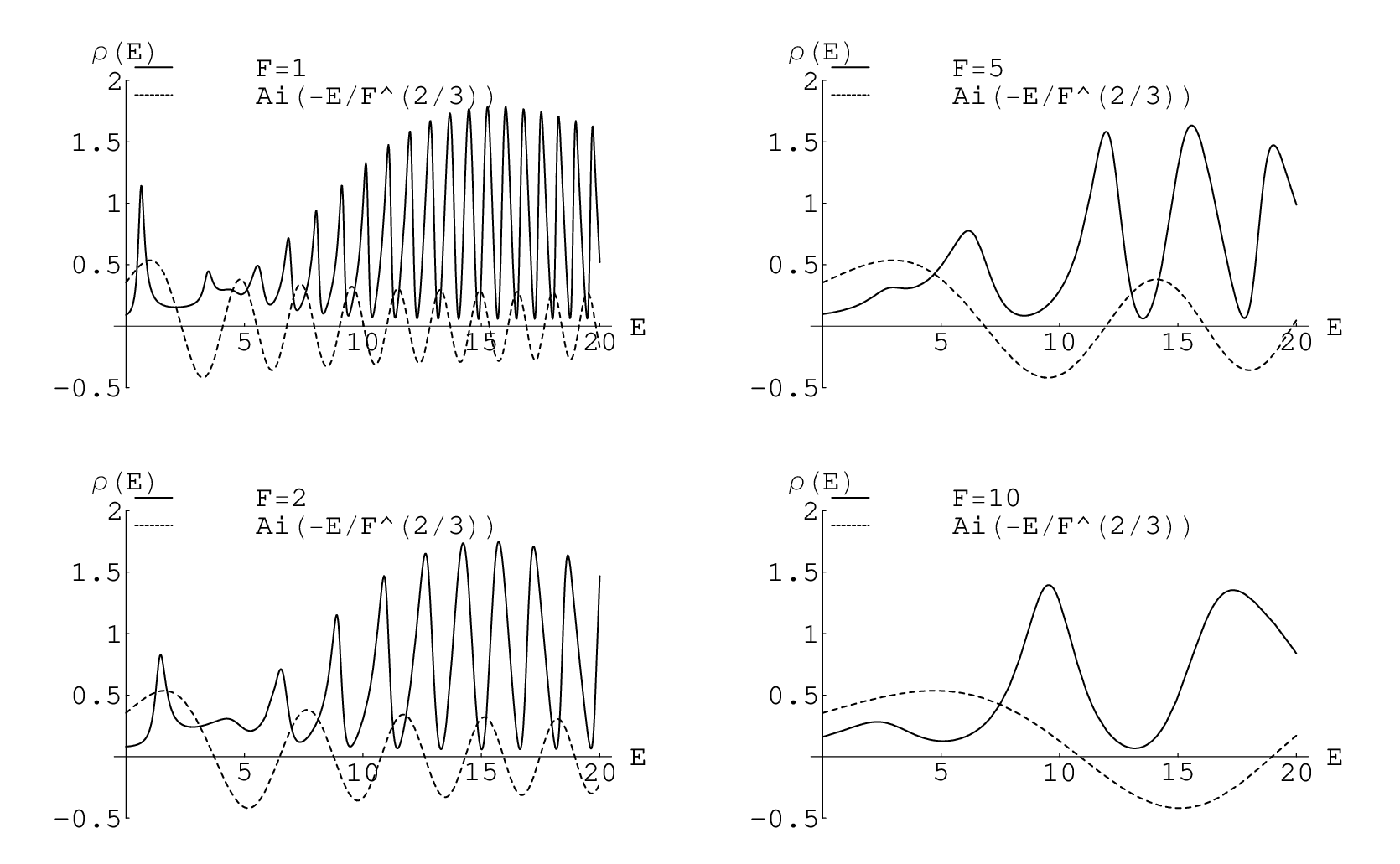}
\caption{Plots of the spectral function (\protect{\ref{starkspectral}}) 
[solid lines] illustrating how the continuum part of the spectrum 
correlates with the zeros of the Airy function $Ai(-E/F^{2/3})$ [dashed 
line], which are the energies of the wedge potential $V=-F x$ with an 
infinite wall at some point. This is an illustration of the Ramsauer 
effect. These plots are for $a=1$ and $g=1.5$, and various values of the 
electric field strength $F$, as shown.}
\end{figure}

\subsubsection{Dependence on the atomic well separation parameter, $a$.}
\begin{figure}[ht]
\includegraphics{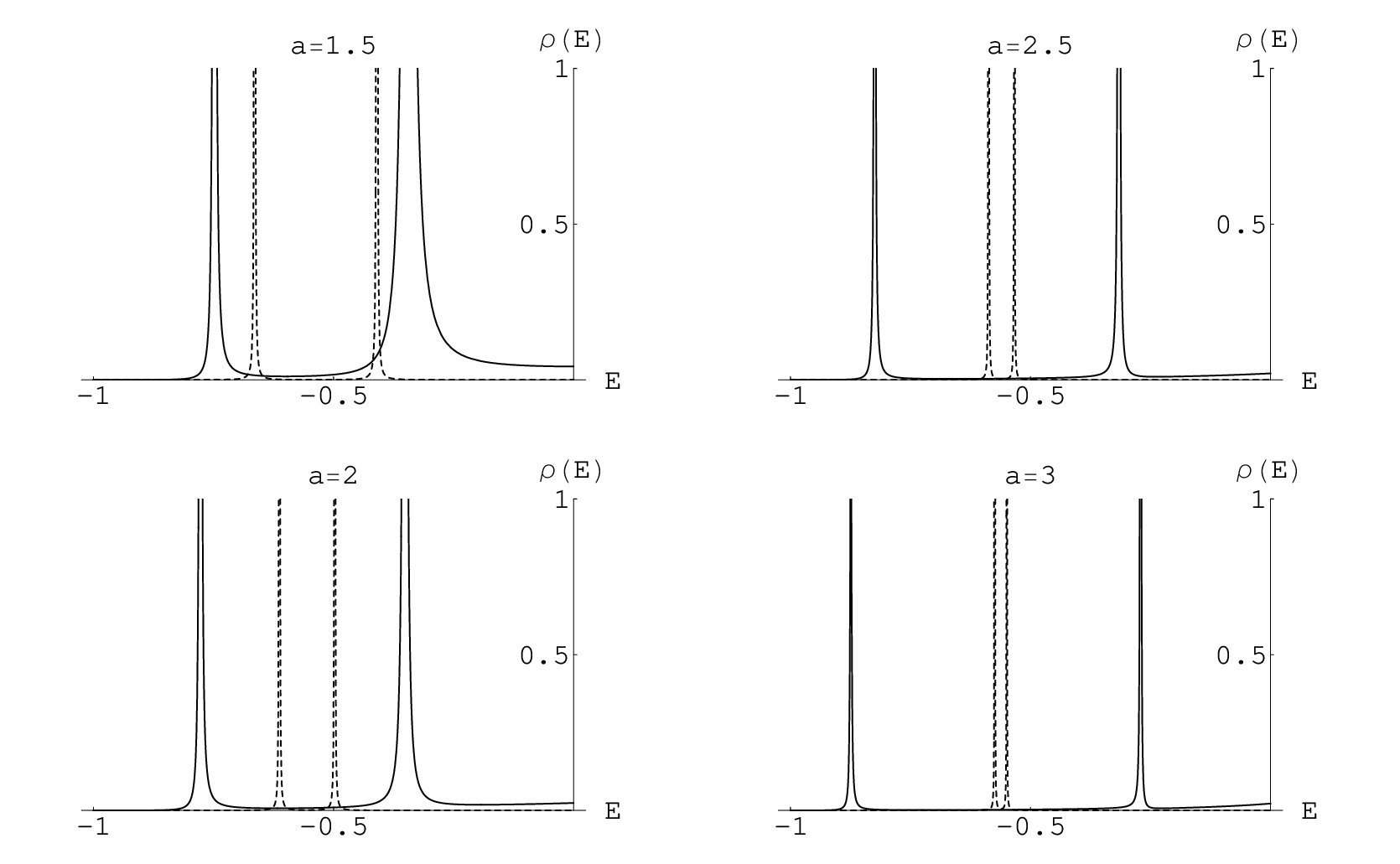}
\caption{Plots of the spectral function (\protect{\ref{starkspectral}}) 
[solid lines] illustrating how the quasi-bound part of the spectrum changes 
as the well separation parameter $a$ varies. The dashed line shows the 
corresponding $F=0$ molecular spectrum (\protect{\ref{molspectral}}). For 
the solid lines, the field strength is $F=0.1$,  the well strength is 
$g=1.5$, and $a$ ranges through $1.5$, $2$, $2.5$ and $3$. Notice that as 
$a$ increases the quasi-bound states become narrower in width and they move 
apart. In contrast, the free bound states move together, eventually becoming degenerate as $a\to\infty$.}
\end{figure}
\begin{figure}[ht]
\includegraphics{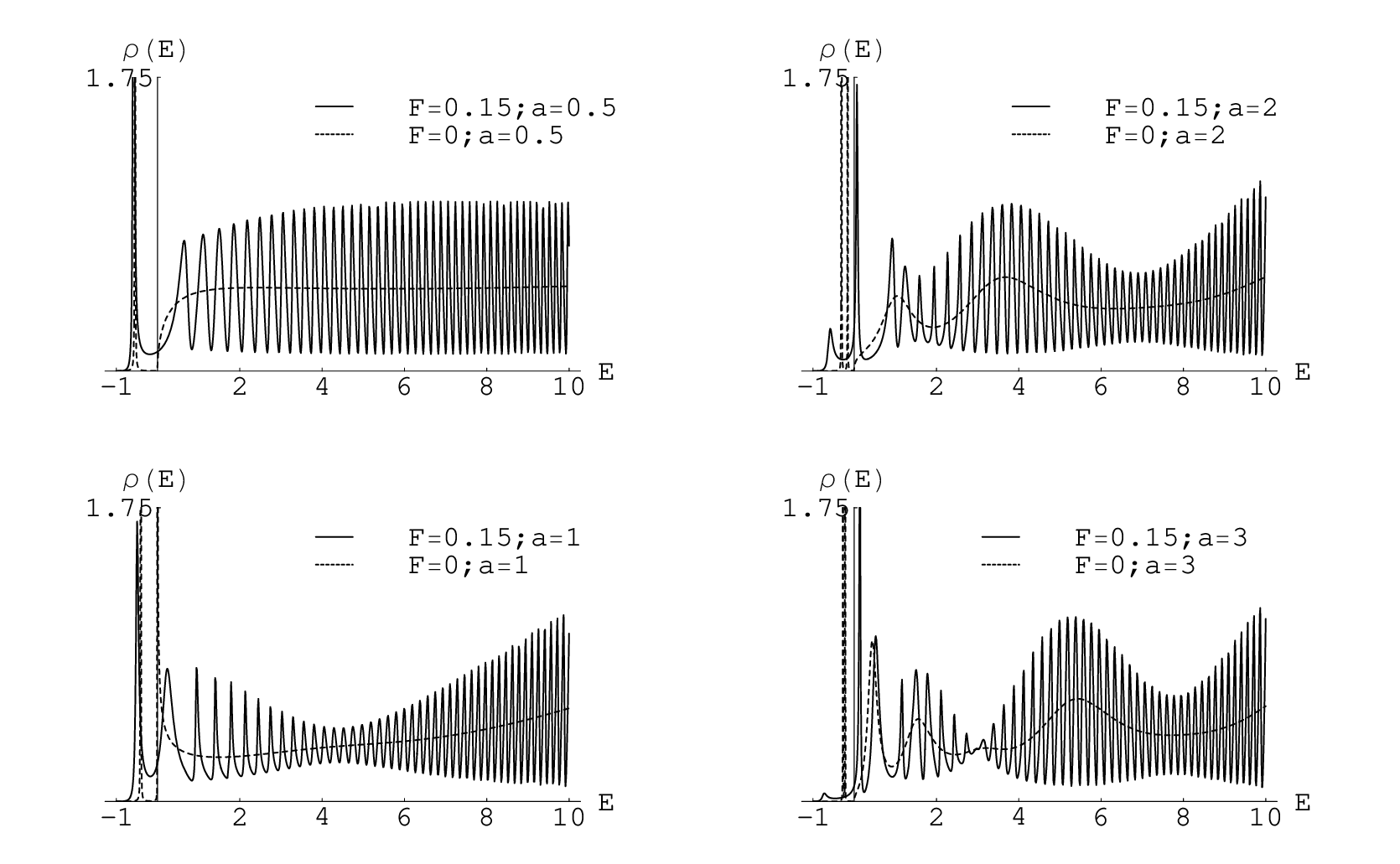}
\caption{Plots of the spectral function (\protect{\ref{starkspectral}}) 
[solid lines] illustrating how the continuum part of the spectrum changes 
as the well separation parameter $a$ varies. The dashed line shows the 
corresponding $F=0$ molecular spectrum (\protect{\ref{molspectral}}). In 
these plots, the field strength is $F=0.15$, and the well strength is 
$g=1$. Notice that as $a$ increases a second quasi-bound state peels off 
the positive energy continuum and forms a Stark-shifted pair around the two 
$F=0$ bound states which exist for $a g >1$. Also note that in the $F=0$ 
case, as $a$ increases the splitting between the two bound states becomes 
vanishingly small, as the tunneling between the two wells is suppressed. In 
the continuum, the average function varies with $a$ due to resonances in 
the backscattering between the two wells. Even with nonzero $F$ the 
spectral function follows this average closely as $a$ varies. }
\end{figure}
The dependence of the spectral function (\ref{starkspectral}) on the atomic 
well separation parameter, $a$, is illustrated in Figures 7 and 8, for the 
"bound" and "continuum" parts of the spectrum, respectively.
In the zero field case the spectral function has two bound states if $g 
a>1$. These are shown as the dashed curves in Figure 7. As the separation 
increases these two bound states approach the same energy, becoming degenerate 
in the limit $a \rightarrow \infty$. This is because in the large separation 
limit the tunneling which mixes the two levels becomes suppressed and the 
two atoms become essentially independent of one another. Thus the bound 
state spectrum approaches that of a single atomic well. On the other hand, 
if the field strength is nonzero the quasi-bound states do not become 
degenerate in the limit of large separation. This is shown by the solid 
curves in Figure 7. Instead, the two states move away from each other; the 
even state is Stark shifted further down in energy and the odd state is 
Stark shifted further up in energy. This is because for large $a$ the 
tunneling is essentially from each well independently, with one Stark 
shifted up and the other down, depending on the parity of the original 
state. These quasi-bound levels also become narrower in width as there is a 
larger barrier through which the electron must tunnel as $a$ increases. The 
effect on the "continuum" part of the spectrum is shown in Figure 8. Here 
$F=0.15$ and the spectrum exhibits resonances which oscillate about the 
 $F=0$ case, which is shown by the dashed lines. As the well 
separation changes, the average function changes, due to the 
Ramsauer-Townsend resonance between the two wells. Note that for nonzero 
field strength the spectral function oscillates rapidly with energy, but 
still follows  the free field average. Thus, the $F>0$ 
spectral function represents a competition between the resonance between the 
two wells, which depends strongly  on $a$, and the resonance between a given 
well and the linear potential, which is not sensitive to $a$.

\subsubsection{Dependence on the atomic well strength, $g$.}
\begin{figure}
\includegraphics{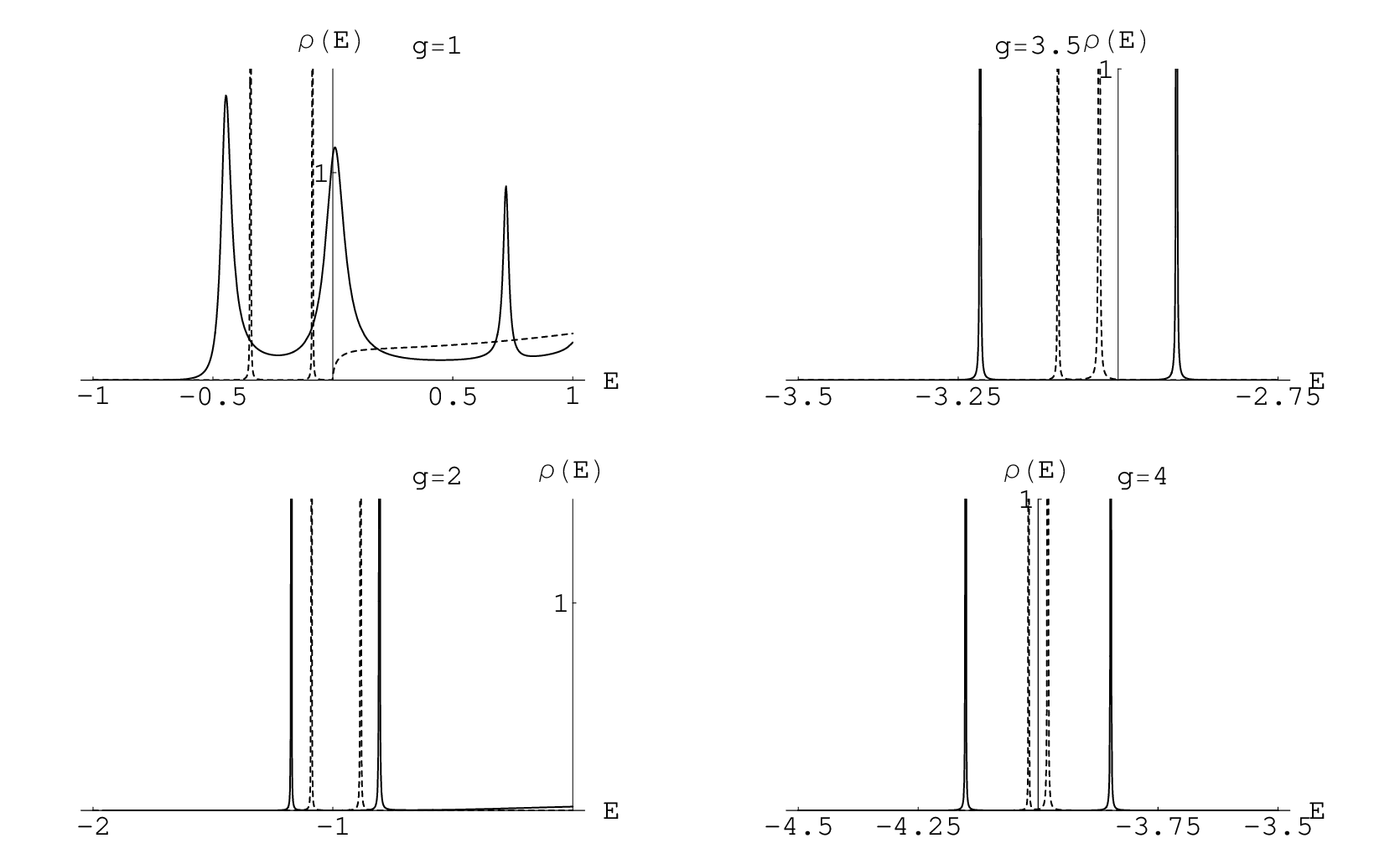}
\caption{Plots of the spectral function (\protect{\ref{starkspectral}}) 
[solid lines] illustrating how the quasi-bound part of the spectrum changes 
as the well depth
parameter $g$ varies. The dashed line shows the corresponding $F=0$ 
molecular spectrum (\protect{\ref{molspectral}}). In these plots, the field 
strength is $F=0.1$, and the well separation parameter is $a=1.5$.}
\end{figure}

The dependence of the spectral function (\ref{starkspectral}) on the atomic 
well depth, $g$, is illustrated in Figures 9 and 10, for the "bound" and 
"continuum" parts of the spectrum, respectively. In the $F=0$ case, the 
difference in energy between the even and odd states gets smaller as the 
well strength increases, which is similar to what happens when the well 
separation increases. This is because the tunneling mixing is more highly 
suppressed as the states become more deeply bound.
The difference between the limit of large well separation and the limit of 
large well strength is that as the well separation increases the two bounded 
states approach the same energy, whereas when the well strength is increased 
the states become degenerate but their energies tend to $-\infty$ as $g 
\rightarrow \infty$. However, if the electric field is applied, the two 
states do not become degenerate in the large well strength limit but instead 
keep a nonzero relative distance between each other. The distance between 
the two states approaches $2Fa$ as well strength increases, which is easily 
explained by the following argument.
If the well strength is very strong we can think  of a particle being 
localized at a single well. If a uniform electric field of strength $F$ 
pointing in the positive x-direction is applied, a particle localized at the 
left well will increase in energy by $Fa$ while the energy of a particle 
localized at the right well will decrease by $Fa$. Since the unperturbed 
states have the same energy the energy difference is just $2Fa$. This can be 
seen in the last panels of Figure 9, for which $F=0.1$ and $a=1.5$, so 
$2Fa=0.3$, which is roughly the separation between the two quasi-bound 
levels. Also note that the quasi-bound state peaks become narrower as $g$ 
increases, as the levels are more deeply bound.

Backscattering resonances in the continuum become more prominent as the well 
strength increases, because the resonances are sharper, since the scattering 
potentials are deeper.  If the field is weak, the WKB approximation for a 
infinite well potential in a electric field can be used to find the energies 
of the backscattering states in the strong well strength limit.  But these 
peaks also follow the free field average (the dashed lines in 
Figure 10), which is due to the  resonance between the wells without the 
electric field.  Thus, the $F>0$ spectral function represents a competition 
between the resonance between the two wells, which depends strongly  on $g$, 
and the resonance between a given well and the linear potential, which is 
also sensitive to the well strength $g$.

\begin{figure}
\includegraphics{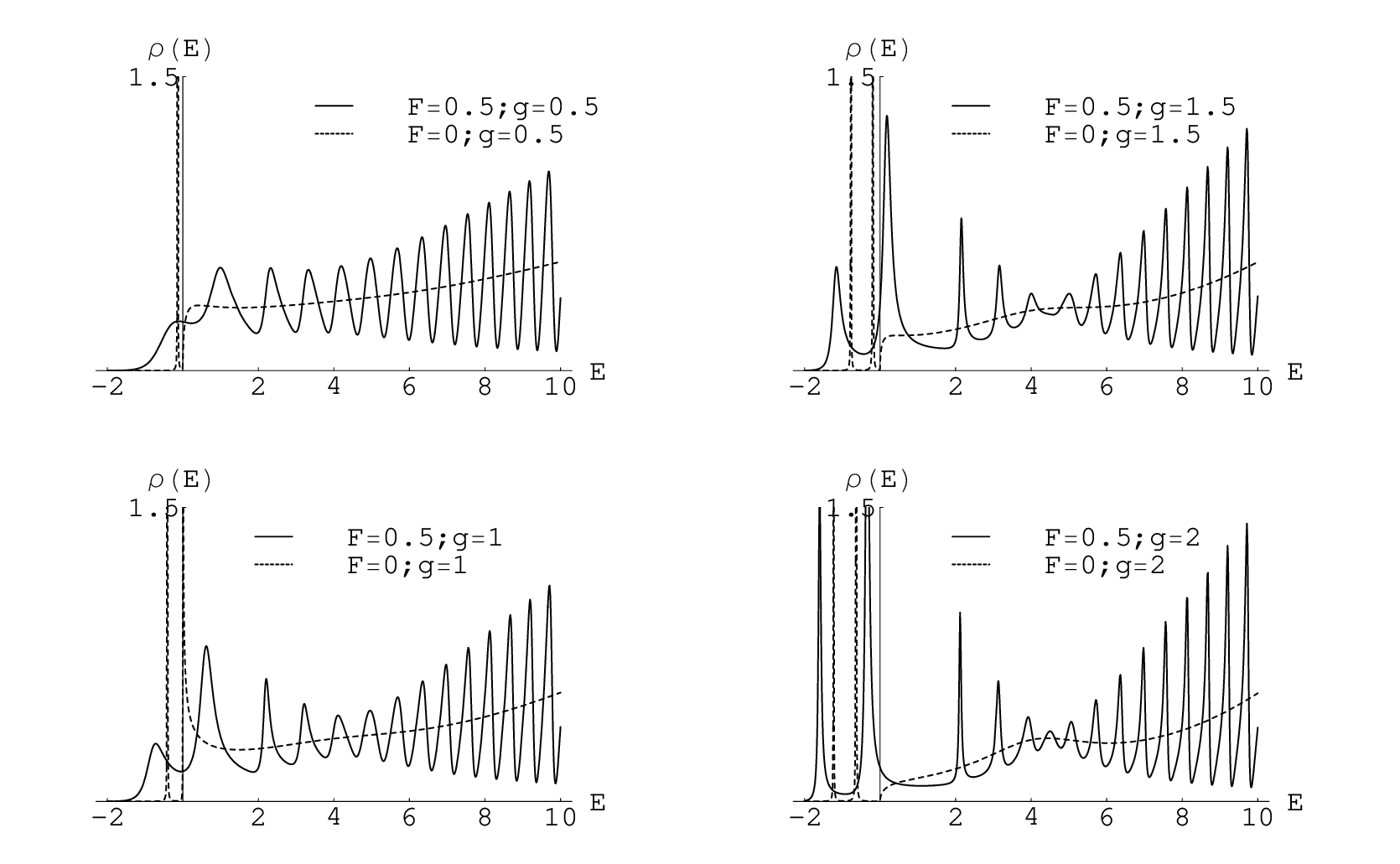}
\caption{Plots of the spectral function (\protect{\ref{starkspectral}}) 
[solid lines] illustrating how the continuum part of the spectrum changes 
as the well depth parameter $g$ varies. The dashed line shows the 
corresponding $F=0$ molecular spectrum (\protect{\ref{molspectral}}). In 
these plots, the field strength is $F=0.5$, and the well separation 
parameter is $a=1$. Notice that as $g$ increases a second quasi-bound state 
peels off the positive energy continuum and forms a Stark-shifted pair 
around the two $F=0$ bound states which exist for $a g >1$.}
\end{figure}

\subsection{Analytic Properties of Spectral Function: Stark Shifts and Level 
Widths.} With the electric field present there are no true bound states, but 
there are quasi-bound states. The location of these quasi-bound states are 
given by the real parts of the poles of the spectral function 
(\ref{starkspectral}), and their widths are given by the imaginary part of 
these poles. These poles are given by the zeros of:
\ba
&&F^{-1/3}+i g^{2}\pi^{2}F^{-1}\Ana\Bigg[\left(\Apa - i \Bpa\right)
\nn\\
&&\times\left(\Apa \Bna - \Ana \Bpa\right)\Bigg]
\nn\\
&&-g \pi F^{-2/3}\Bigg[i\Ana^{2}+\Ana \Bna
\nn\\
&&+i\Apa^{2}+\Apa \Bpa\Bigg]
\label{poles}
\ea
In the weak field limit we can use the following asymptotic expansions 
\cite{abram} of the Airy functions to find perturbative solutions for the 
zeros of (\ref{poles}):
\ba
Ai(z)&\sim& \frac{e^{-\zeta}}{2\sqrt{\pi} z^{1/4}}\,\sum_{k=0}^\infty  
\frac{(-1)^k\, c_k}{\zeta^k}
\label{airyaasymptotic}\\
Bi(z)&\sim& \frac{e^{\zeta}}{\sqrt{\pi} z^{1/4}}\,\sum_{k=0}^\infty 
\frac{c_k}{\zeta^k}
\label{airybasymptotic}
\ea
where $\zeta=\frac{2}{3}z^{3/2}$, and the expansion coefficients $c_k$ are
\ba
c_k=\frac{\Gamma(3k+1/2)}{54^k\, k!\, \Gamma(k+1/2)}
\label{cs}
\ea
Clearly, only even powers of $F$ will appear in the perturbative expansion 
for the real parts of the quasi-energies. So, we define the expansion
\ba
E=-g^2\sum_{n=0}^\infty a_n \left(\frac{F}{g^3}\right)^{2n}
\label{ptenergy}
\ea
To find the real parts of the quasi-energies, we can ignore the imaginary 
parts of (\ref{poles}), which are anyway exponentially suppressed in the 
weak field limit. It is a straightforward exercise to expand (\ref{poles}) 
in powers of the field strength $F$, using for example {\it Mathematica} 
\cite{wolfram}. The leading order $F^0$ term produces the equation
\ba
4a_0-4\sqrt{a_0}+\left(1-e^{-4 a g \sqrt{a_0}}\right)=0
\label{lead}
\ea
whose solutions are just the solutions of the transcendental equations 
(\ref{moleven}) and (\ref{molodd}) derived in the previous section for the 
bound  states in the free field case. A solution to (\ref{lead}) satisfying 
(\ref{moleven}) is an even bound state of the $F=0$ potential 
(\ref{potential}), while a solution to (\ref{lead}) satisfying 
(\ref{molodd}) is an odd bound state of (\ref{potential}).

The first correction to these bound states comes from the $F^2$ term in the 
expansion of (\ref{poles}), which leads to the following expression for 
$a_1$ in terms of $a_0$:
\ba
a_1=\frac{15(1-2\sqrt{a_0})-15 a g (1-2\sqrt{a_0})^2+12 (a g)^2 
a_0(1-4\sqrt{a_0})-4(a g)^3 a_0(1-2\sqrt{a_0})^2}{48a_0^2 
(1-2\sqrt{a_0})(1-a g(1-2\sqrt{a_0}))}
\label{next}
\ea
Thus, to first nontrivial order, the Stark shifted energy is
\ba
E=-a_0 g^2  -a_1 \frac{F^2}{g^4} + \dots
\ea
To find the shift for the even bound state, we find the solution $a_0$ of 
(\ref{leading}) which also satisfies (\ref{moleven}), and then insert this 
value of $a_0$ into (\ref{next}) to find the corresponding $a_1$. It is 
straightforward to continue this to higher orders. (We study details of the 
higher orders in the Section \ref{atomelectric} for the special "atomic" 
case where the atomic separation parameter $a=0$.) For the odd bound state, we must first find if there is such a solution for $a_0$ to (\ref{leading}) satisfying 
(\ref{molodd}). This odd solution will exist if $g a>1$. If it exists, then 
the corresponding Stark shift is obtained by inserting this value of $a_0$ 
into the expression (\ref{next}) for $a_1$.

The widths of the quasi-bound states can be derived from the imaginary part 
of the poles of the spectral function.  Because of the complicated dependence of 
the quasi-bound state energies on the system parameters $F$, $g$ and $a$, 
it is difficult to derive simple analytical expressions for the line widths. However, in the large atomic separation limit, as $a\to\infty$, we can use WKB to approximate the tunneling rate, and hence the line width, as
\begin{equation}
\Gamma \sim g\sqrt{E_{0}-Fa}\exp\left[-\frac{4 (E_{0}-F 
a)^{3/2}}{3 F}\right]
\label{width}
\end{equation}
where $E =-E_{0}-i \Gamma$, and $-E_{0}$ is the full Stark-shifted energy of the 
lower quasi-bound state.

In the limit of infinitely large well strength, $g\to\infty$, the $g^{2}$ term in 
(\ref{poles}) dominates and the zeros of (\ref{poles}) lie on the positive real 
axis. If the field strength $F$ is small, the first order asymptotic expansion of the 
Airy functions \cite{abram}:
\begin{eqnarray}
Ai(-z)
&\sim&
\frac{1}{\sqrt{\pi} z^{1/4}}\sin \left(\zeta+\frac{\pi}{4} \right)
\label{alarge}\\
Bi(-z)
&\sim&
\frac{1}{\sqrt{\pi} z^{1/4}}\cos \left(\zeta+\frac{\pi}{4} \right)\qquad ; \quad \zeta=\frac{2}{3}z^{3/2}
\label{blarge}
\end{eqnarray}
for $Re(z)\gg 0$ can be used to estimate the location of the zeros of 
(\ref{poles}). These
approximate zeros are determined by the expressions:
\begin{eqnarray}
\frac{2}{3}\frac{(E-Fa)^{3/2}}{F}
&=&\left(n-\frac{1}{4}\right) \pi
\label{FWRExpression}\\
\frac{2}{3}\frac{(E+Fa)^{3/2}-(E-Fa)^{3/2}}{F}&=&n\, \pi
\label{WWRExpression}
\end{eqnarray}
Note that expression (\ref{FWRExpression}) approximates the zeros of $Ai
\left(-\frac{(E-Fa)}{F^{2/3}} \right)$, while (\ref{WWRExpression})
approximates the zeros of $Ai\left(-\frac{(E+Fa)}{F^{2/3}} \right)Bi 
\left(-\frac{(E-Fa)}{F^{2/3}} \right)-Ai
\left(-\frac{(E-Fa)}{F^{2/3}} \right)Bi \left(-\frac{(E+Fa)}{F^{2/3}} 
\right)$.
Solutions for expression (\ref{FWRExpression}) are the 
energies obtained using the WKB approximation for a half wedge potential 
with an infinitely high wall at
$x=-a$. Therefore, we interpret the energies satisfying (\ref{FWRExpression}) 
as those of the backscattering states between the left well and the electric 
field, which is the Ramsauer effect illustrated in Figures 5 and 6. Similarly,
expression (\ref{WWRExpression}) yields the same set of energies obtained from 
the WKB approximation for a potential well in an electric field with infinitely high 
walls at $x=-a$ and $x=a$, and therefore (\ref{WWRExpression}) determines the energies of backscattering states between the two wells, also in the presence of the electric field. In general, the exact resonances reflect a competition of these scatterings amongst the linear potential and the atomic wells.

\section{Single Well Potential With an Electric Field}

\label{atomelectric}

A nice feature of our molecular analysis is that the "atomic" analogue of the molecular model studied in Section 
\ref{molelectric} can be obtained simply by setting the separation 
parameter, $a$, of the two wells to zero. All the expressions carry over 
smoothly in this $a\to 0$ limit. The corresponding potential is (note that 
the delta function strength becomes $2g$ in this limit):
\ba
V(x)=-2 g \, \delta(x)-F\,x
\label{atompotential}
\ea
This atomic problem has also been discussed in terms of the 
corresponding Green's functions in \cite{ludviksson,soldati}.

\subsection{WTK Solution for the Atomic Spectral Function}
The spectral function is given, as before in (\ref{starkspectral}), by
\ba
\rho(E)=\lim_{\epsilon \rightarrow 0}
\frac{1}{\pi} Im
\textstyle\left(\frac{B_{-}^{(u)}(E+i\epsilon)\left[B_{+}^{(u)}(E+i\epsilon)+iA_{+}^{(u)}(E+i\epsilon)\right]+
B_{-}^{(v)}(E+i\epsilon)\left[B_{+}^{(v)}(E+i\epsilon)+iA_{+}^{(v)}(E+i\epsilon)\right]}
{B_{-}^{(u)}(E+i\epsilon)\left[B_{+}^{(v)}(E+i\epsilon)+iA_{+}^{(v)}(E+i\epsilon)\right]-
B_{-}^{(v)}(E+i\epsilon)\left[B_{+}^{(u)}(E+i\epsilon)+iA_{+}^{(u)}(E+i\epsilon)\right]}\right)
\label{atomspectral}
\ea
where the coefficient functions in (\ref{pluscoeff}) and (\ref{minuscoeff}) now simplify to :
\ba
A_{+}^{(u)}(E)&=&\pi \left(
Bi^\prime\left(-\frac{E}{F^{2/3}}\right) - g F^{-1/3} 
Bi\left(-\frac{E}{F^{2/3}}\right)\right)
\nn\\
B_{+}^{(u)}(E)&=&-\pi \left(
Ai^\prime\left(-\frac{E}{F^{2/3}}\right)
-g F^{-1/3} Ai\left(-\frac{E}{F^{2/3}}\right)\right)
\nn\\
A_{+}^{(v)}(E)&=&-\pi F^{-1/3} Bi\left(-\frac{E}{F^{2/3}}\right)\nn\\
B_{+}^{(v)}(E)&=&\pi F^{-1/3} Ai\left(-\frac{E}{F^{2/3}}\right)\nn\\
A_{-}^{(u)}(E)&=&\pi \left( Bi^\prime\left(-\frac{E}{F^{2/3}}\right) + g 
F^{-1/3} Bi\left(-\frac{E}{F^{2/3}}\right)\right)\nn\\
B_{-}^{(u)}(E)&=&-\pi \left( Ai^\prime\left(-\frac{E}{F^{2/3}}\right)
+g F^{-1/3} Ai\left(-\frac{E}{F^{2/3}}\right)
\right)\nn\\
A_{-}^{(v)}(E)&=&-\pi F^{-1/3} Bi\left(-\frac{E}{F^{2/3}}\right)\nn\\
B_{-}^{(v)}(E)&=&\pi F^{-1/3} Ai\left(-\frac{E}{F^{2/3}}\right)
\label{atomcoeff}
\ea
Taylor expanding in $\epsilon$ leads to a more explicit expression for the 
spectral function :
\ba
\rho(E)&=&\frac{1}{\left[2 \pi  g F^{-2/3} \A^{2}\right]^{2}+\left[2 \pi g 
F^{-2/3} \B \A - F^{-1/3}\right]^{2}}
\nn\\
&&\hskip -2cm
\times
\bigg(\Big[
( (1-g^{2}) \pi F^{-2/3} \A \B + \pi \dA \dB + g\,F^{-1/3})
\nn\\
&&\times
( 2 g F^{-2/3} \A^{2})
\Big]
-\Big[
(2 \pi g F^{-2/3} \B \A - F^{-1/3})
\nn\\
&&
\times
((1-g^{2})F^{-2/3}
\A^{2}+\dA^{2})
\Big]
\bigg)
\ea
\subsection{Large-Order Perturbation Theory and Level Widths.}

It is well known that perturbation theory for an unstable quasi-bound state 
produces a divergent non-alternating series, with which an imaginary part 
can be associated using Borel techniques. This was first explicitly 
investigated for the unstable $x^2+g x^3$ anharmonic potential 
\cite{arkady}, then in extensive detail for the $x^2+\lambda x^4$ potential 
\cite{benderwu}, and is a very general property of perturbation theory 
\cite{zinn,gdreview}. The same ideas apply to the Stark effect problem, as 
has been investigated exhaustively for atomic systems
\cite{silverstone,borel,ivanov,jentschura}.

The quasi-bound states correspond to poles of the spectral function, namely 
solutions to
\ba
\left[2 \pi  g F^{-2/3} \A^{2}\right]^{2}+\left[2 \pi g F^{-2/3} \B \A - 
F^{-1/3}\right]^{2}=0
\label{atompoles}
\ea
The real part of the quasi-energy can be found by making a perturbative 
expansion for the real part, $E_{\rm real}$, of the energy
\ba
E_{\rm real}=-g^2\sum_{n=0}^\infty a_n \left(\frac{F}{g^3}\right)^{2n}
\label{atompt}
\ea
as in (\ref{ptenergy}). When the field strength $F$ vanishes, there is a 
single (even) bound state at $E=-g^2$. Thus, $a_0=1$, as is consistent with 
(\ref{lead}) when $a\to 0$. The width of the quasi-bound state can be 
estimated by writing
\ba
E=E_{\rm real} + i E_{\rm imag}
\ea
where we expect $E_{\rm imag}$ to be exponentially small. Indeed, expanding 
the imaginary part of (\ref{atompoles}) immediately leads to the leading 
behavior
\ba
E_{\rm imag}\sim - g^2\, \exp\left[-\frac{4g^3}{3F}\right]
\label{atomimag}
\ea
in agreement with the $a\to 0$ limit of the molecular case (\ref{width}).

We now show how this is consistent with an analysis of the divergence of the 
perturbative expansion for the real part of the quasi-bound energy level. 
First, note that the $Ai^2$ term in (\ref{atompoles}) is exponentially small 
in the small $F$ limit, and so can be neglected. Thus the real part is 
determined by
\ba
Ai\left(\frac{g^2}{F^{2/3}}\sum_{n=0}^\infty a_n 
\left(\frac{F}{g^3}\right)^{2n}\right) 
Bi\left(\frac{g^2}{F^{2/3}}\sum_{n=0}^\infty a_n 
\left(\frac{F}{g^3}\right)^{2n}\right)
=\frac{F^{1/3}}{\pi g}
\label{atomcondition}
\ea
In the small $F$ limit we can use the asymptotic expansions 
(\ref{airyaasymptotic},\ref{airybasymptotic}) of the Airy functions to make an expansion of 
(\ref{atomcondition})  in powers of $(F/g^3)^2$, thereby successively determining the 
coefficients $a_n$. It is trivial to program this expansion in 
{\it Mathematica}, and the results for the first 21 expansion coefficients 
are shown in Table 1 \cite{roman}.

\begin{table}
\begin{tabular}[h]{|c|c|c|c|c|}
\hline
$n$ & $a_{n}$ & $N[a_n]$ & $N \left[ a_n/a_n^{\rm (lead)} \right]$ & $R\left[ a_n/a_n^{\rm (lead)} \right]$ \\
\hline
$0$ & $1$ & $1$ & - & - \\
$1$ & $\frac{5}{16}$ & $0.3125$ & $0.872665$ & $1.04760$ \\
$2$&$\frac{55}{64}$&$0.859$&$0.71106$&$0.95104$\\$
3$&$\frac{10625}{1024}$&$10.376$&$0.763133$&$0.9533$\\$
4$&$\frac{1078125}{4096}$&$263.214$&$0.819423$&$0.98733$\\$
5$&$\frac{366940625}{32768}$&$11198.1$&$0.860776$&$1.006199$\\$
6$&$\frac{93784578125}{131072}$&$715520.$&$0.888895$&$1.008710$\\$
7$&$\frac{269028257953125}{4194304}$&$6.4 \, .\,  10^7$ & $0.908071$ & $1.005397$\\$
8$&$\frac{129011616275390625}{16777216}$&$7.7 \, .\,  
{10}^9$&$0.921614$&$1.002243$\\$
9$&$\frac{159621687625662109375}{134217728}$&$1.2 \, .\,  
{10}^{12}$&$0.931602$&$1.000556$\\$
10$&$\frac{123839968932138228515625}{536870912}$&$2.3 \, .\,  
{10}^{14}$&$0.939271$&$0.99992$\\$
11$&$\frac{471147487418797446943359375}{8589934592}$&$5.5 \, .\,  
{10}^{16}$&$0.945356$&$0.99978$\\$
12$&$\frac{539212883805702339810673828125}{34359738368}$&$1.6 \, .\,  
{10}^{19}$&$0.950311$&$0.99981$\\$
13$&$\frac{1462185114846262625626556396484375}{274877906944}$&$5.3 \, .\,  
{10}^{21}$&$0.95443$&$0.99987$\\$
14$&$\frac{144871600275431039774199176025390625}{68719476736}$&$2.1 \, .\,  
{10}^{24}$&$0.957911$&$0.99991$\\$
15$&$\frac{67969184060037298421788742225469970703125}{70368744177664}$&$9.7 
\, .\,  {10}^{26}$&$0.960894$&$0.99995$\\$
16$&$\frac{142608185435906164633493702703533111572265625}{281474976710656}$&$5.1\, 
.\,  {10}^{29}$&$0.963479$&$0.99997$\\$
17$&$\frac{679265718819054465192747030828993319061279296875}{2251799813685248}$&$3.0 
\, .\,  {10}^{32}$&$0.965742$&$-$\\$
18$&$\frac{1822495852683481842017384269925359639728546142578125}{9007199254740992}$&$2.0 
\, .\,  {10}^{35}$&$0.96774$&$-$\\$
19$&$\frac{21888188031753229357462565895827650045023616790771484375}{144115188075855872}$&$1.5 
\, .\,  {10}^{38}$&$0.969516$&$-$\\$
20$&$\frac{73105891881984796538857909985635411709301443950653076171875}{576460752303423488}$&$1.3 
\, .\,  {10}^{41}$&$0.971107$&$-$\\
\hline
\end{tabular}
\caption{This table lists the coefficients $a_n$ appearing in the 
perturbative expansion (\protect{\ref{atompt}}) for the quasi-energy level 
fo the atomic system in an electric field. Note that these coefficients are 
non-alternating in sign, and that their magnitude, $N[a_n]$, grows very fast 
with the perturbative order  $n$, as is shown in the third column. The 
fourth column shows the ratio of the $a_n$ to the leading factorial growth 
rate in (\protect{\ref{leading}}), and the fifth column gives the 
fourth-order Richardson extrapolation \cite{carlbook} of this ratio, showing 
its rapid approach to unity.}
\end{table}

Notice that all the $a_n$ have the same sign, and their magnitude grows very 
rapidly. In fact, the leading growth rate at large orders $n$ of the 
perturbation series is factorially fast:
\ba
a_n^{(lead)}=\frac{2}{\pi}\left(\frac{3}{4}\right)^{2n}\Gamma(2n)
\label{leading}
\ea
The last two columns of Table 1 show how quickly this leading growth rate is 
achieved. The final column uses fourth order Richardson extrapolation 
\cite{carlbook} for the ratio of the exact coefficients to the leading 
behavior in (\ref{leading}), and the rapid onset to the leading growth shows 
that after 10 terms one is already well and truly into the asymptotic regime 
described by (\ref{leading}).

Since the perturbative series (\ref{atompt}) for the energy is divergent and 
non-alternating, it is not Borel summable \cite{carlbook}. However, one can 
associate an imaginary part with such a divergent series using standard 
Borel dispersion relations \cite{arkady,benderwu,zinn,gdreview}. This technique provides a 
bridge between perturbative and nonperturbative physics, and is well studied 
in a wide variety of physics contexts \cite{zinn}.
The argument  can be motivated and illustrated roughly as follows. Consider the
following {\it alternating} divergent series obtained by an asymptotic 
expansion of the integral
\ba
\int_0^\infty \frac{dt}{t}\, \frac{e^{-t/c}}{1+t^2}\sim \sum_{n=0}^\infty 
(-1)^n c^{2n}\Gamma(2n)
\label{borel}
\ea
where $c>0$. The integral on the left-hand-side of (\ref{borel}) is the 
Borel sum of the alternating divergent series of the right-hand-side. This 
argument does not work for the corresponding nonalternating series, 
$\sum_{n=0}^\infty  c^{2n}\Gamma(2n)$, as one would encounter a pole on the 
$t$ integration axis. The behavior at such a pole must be defined by some sort 
of prescription. The standard principal parts prescription associates with 
this non-alternating divergent series a non-perturbative imaginary part 
\cite{zinn,gdreview}
\ba
{\it Im}\left(\sum_{n=0}^\infty  c^{2n}\Gamma(2n)\right) \sim 
\frac{\pi}{2}\, \exp\left[-\frac{1}{c}\right]
\label{pp}
\ea
Thus, using the leading growth rate (\ref{leading}) for the perturbative 
coefficients $a_n$ we see that this Borel approach associates an imaginary 
part for the quasi-bound energy level equal to
\ba
{\it Im}(E)\sim -g^2 \,\exp\left[-\frac{4g^3}{3 F}\right]
\ea
This agrees precisely with the imaginary part (\ref{atomimag}) found 
directly from the properties of the Airy functions, as well as with a simple 
WKB estimate.

\section{Conclusions}

In this paper we have presented the exact analytic solution for the spectral 
function for the simple one-dimensional molecular ionization model of a 
diatomic molecule represented by two attractive delta function wells in an 
external static electric field. The Weyl-Titchmarsh-Kodaira spectral theorem 
provides a simple construction for the spectral function in terms of 
suitably normalized solutions to the Schr\"odinger equation. In this case 
these solutions are Airy functions, and the spectral function can be 
expressed in closed form in terms of Airy functions. Thus, the spectral 
function can easily be plotted using a program such as {\it Mathematica} 
\cite{wolfram}. The dependence of the spectral function on the relevant 
physical parameters, the field strength $F$, the well strength $g$, and the 
well separation parameter $a$, is illustrated in Section 4 by a collection 
of plots. This helps develop a body of intuition for the behavior of the 
quasi-bound states as they are Stark shifted and broadened, and also for the 
resonance structures in the "continuum", which reflect a competition between 
Ramsauer-Townsend resonant scattering between the two atomic wells and 
between one or both atomic well(s) and the linear field potential. The most 
important extension of this model would be to consider the effect of 
time-dependence in the background electric field, which introduces yet 
another physical scale into the problem 
\cite{keldysh,geltman2,reiss,chu,starace}.

{\bf Acknowledgement:} We thank George Gibson, Misha Ivanov and Roman Jackiw 
for helpful comments and suggestions. We also thank Stephen Fulling and Larry Glasser for 
bringing references \cite{ludviksson} and \cite{glasser} to our attention.

\end{document}